\documentclass[%
 reprint,longbibliography,preprintnumbers,
nofootinbib,
 amsmath,amssymb,
 aps
]{revtex4-2}
\pdfoutput=1
\usepackage{graphicx}
\usepackage[utf8]{inputenc}
\usepackage{flushend}
\usepackage{dcolumn}
\usepackage{bm}
\usepackage{balance}

\usepackage[normalem]{ulem}

\usepackage[colorlinks = true,
            linkcolor = blue,
            urlcolor  = blue,
            citecolor =green,
            anchorcolor = blue]{hyperref}
\usepackage{verbatim}
\usepackage{color,ulem}
\usepackage[english]{babel}

\usepackage[utf8]{inputenc}
\input Starburst.fd
\newcommand*\initfamily{\usefont{U}{Starburst}{xl}{n}}\initfamily

\newcommand{\beq}{\begin{eqnarray}}
\newcommand{\eeq}{\end{eqnarray}}
\usepackage{amsmath}
\usepackage{tikz}
\usetikzlibrary{decorations.pathmorphing}
\usetikzlibrary{shapes.misc}
\tikzset{cross/.style={cross out, draw=black, minimum size=8*(#1-\pgflinewidth), inner sep=0pt, outer sep=0pt},
cross/.default={1pt}}
\usetikzlibrary{patterns,math}
\begin{document}

\title{\Large  Ultrastable 2D glasses and packings explained by local centrosymmetry }

\author{\textbf{Alessio Zaccone}$^{1}$}%
 \email{alessio.zaccone@unimi.it}
 
 \vspace{1cm}

\affiliation{$^{1}$Department of Physics ``A. Pontremoli'', University of Milan, via Celoria 16,
20133 Milan, Italy.}

\begin{abstract}
Using the most recent numerical data by Bolton-Lum \emph{et al.} [Phys. Rev. Lett. 136, 058201 (2026)], we demonstrate that ideal ultrastable glasses in the athermal limit (or ultrastable ideal 2D disk packings) possess a remarkably high degree of local centrosymmetry. In particular, we find that the inversion-symmetry order parameter for local force transmission introduced in Milkus and Zaccone, [Phys. Rev. 93, 094204 (2016)], is as high as $F_{IS}= 0.84522$, to be compared with $F_{IS}=1$ for perfect centrosymmetric crystals free of defects, and with $F_{IS}\sim 0.3-0.5$ for standard random packings. This observation provides a clear, natural explanation for the ultra-high shear modulus of ideal packings and ideal glasses, because the high centrosymmetry prevents non-affine relaxations which decrease the shear modulus. The same mechanism explains the absence of boson peak-like soft vibrational modes. These results also confirm what was found previous work, i.e. that the bond-orientational order parameter is a very poor correlator for the vibrational and mechanical properties of disordered packings.
\end{abstract}

\maketitle
Ideal or ultrastable glasses have emerged as a useful groundwork to study long-standing open issues such as the nature of the glass transition. Understanding their properties also bears relevance for experimental systems, where ultrastable glasses have been realized since their discovery by Ediger and co-workers in 2007 using vapor deposition techniques \cite{Ediger}.
Yet, there are still a lot of open questions, and often misconceptions, about the origin of the unusual properties of ultrastable glasses, which appear to be more similar to those of crystals \cite{Taiki,McKenna}. It thus appears like a paradox, that these systems have mechanical and vibrational properties akin to crystals, but their structure is completely disordered and random, as reflected in low values of traditional metrics such as the bond-orientational order parameter.

\begin{figure}
    \centering
    \includegraphics[width=\linewidth]{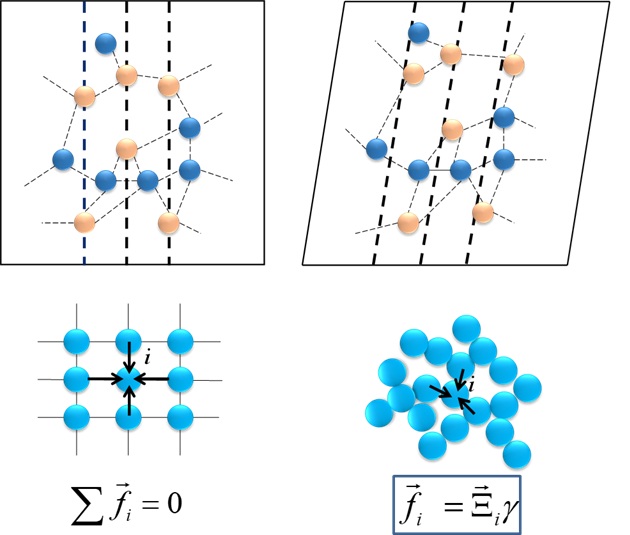}
   \caption{Schematic illustration of the connection between local inversion symmetry and nonaffine displacements in disordered solids. \textit{Top:} An imposed shear deformation defines an affine displacement field (dashed lines). In inversion--asymmetric environments particles experience a net force and undergo nonaffine displacements away from the dashed lines. \textit{Bottom:} In centrosymmetric environments the sum of forces vanishes, while inversion-symmetry breaking induces a finite force imbalance in the affine position that drives nonaffine rearrangements. This force--imbalance mechanism underlies nonaffinity in disordered solids \cite{Zaccone2011,Lemaitre,Zaccone_book}.}
    \label{fig:1}
\end{figure}

In this paper, we provide the missing geometric link without which it is impossible to resolve the above paradox.
By analyzing the ultrastable 2D random disk packings of Ref. \cite{Corwin}, we compute a metric which captures the local statistical degree of inversion-symmetry around each disk. This metric directly connects to the microscopic origin of nonaffine displacements in the deformation and lattice dynamics of amorphous materials. The core concept is easily explained: in a perfect centrosymmetric crystal subjected to a shear deformation, each particle will receive forces from its nearest-neighbours which cancel each other out by inversion-symmetry. Hence, each particle is already at mechanical equilibrium in the (affine) position prescribed by the deformation tensor. Conversely, in a disordered lattice with no centrosymmetry, under strain, each particle will receive forces from its nearest-neighbors, which do not cancel each other out by symmetry, thus leaving a net unbalanced force in the affine position \cite{Zaccone_book,Lemaitre, Schlegel2016, JAP,ZacconePRL}. This net force, in turn, triggers an additional (nonaffine) displacement, on top of the affine one dictated by the applied strain. The nonaffine displacements occur in the local force field of interparticle interactions and, because force times displacement is a work, specifically an internal work done by the particles to keep mechanical equilibrium, this contributes negatively to the internal energy of deformation and leads to a negative contribution to the shear modulus.
The mechanism is schematically illustrated in Fig. \ref{fig:1}.

For isostatic packings, it has been shown rigorously \cite{Zaccone2011} that this negative nonaffine contribution exactly cancels the affine contribution, thus leading to the vanishing of rigidity or unjamming at $z_c=2d$ for central-force packings at the random close packing (RCP) \cite{Zaccone2022,Zaccone_rev}, where $z$ is the coordination number and $d$ the space dimension. 
Therefore, the statistical degree of centrosymmetry in local force-transmission, introduced in Ref. \cite{Milkus2016}, can be leveraged as a useful metric to predict the mechanical and vibrational properties of a amorphous systems solely based on the static particle-level structure of the system, and this usefulness has been benchmarked several times in the past on experimental data for very different systems such as colloidal glasses \cite{Amelia,Amelia2}, jammed packings \cite{Wang2025} and metallic glasses \cite{entropy}.
Furthermore, it was shown already in \cite{Milkus2016} that the inversion-symmetry order parameter $F_{IS}$ not only provides a strong correlation with the mechanical response but also with the vibrational properties, such as the boson peak excess of soft vibrational modes over the Debye prediction. At the same time, a traditional metric such as the bond-orientational order parameter \cite{ronchetti} is unable to correlate with the mechanical and vibrational properties of disordered systems such as random packings and random networks \cite{Milkus2016}.

Bolton--Lum \emph{et al.}~\cite{Corwin} recently introduced numerically generated
two-dimensional polydisperse disk packings that exhibit several striking properties,
including hyperuniformity \cite{torquato}, a finite shear modulus in the low-pressure limit, and
a vibrational spectrum lacking boson-peak–like excess modes. The latter two features are
unusual for standard disordered packings \cite{OHern,torquato_rev} and suggest a predominantly affine elastic
response. Furthermore, the packing fraction found in \cite{Corwin} of about $0.91$ is significantly lower than the maximal packing fractions for random disk packing of polydisperse disks which may exceed $0.96$ for very wide size distributions \cite{Zaccone2025_RCP2D}.
The structural origin of these crystal-like properties, however, remains unclear.

\begin{figure}
    \centering
    \includegraphics[width=\linewidth]{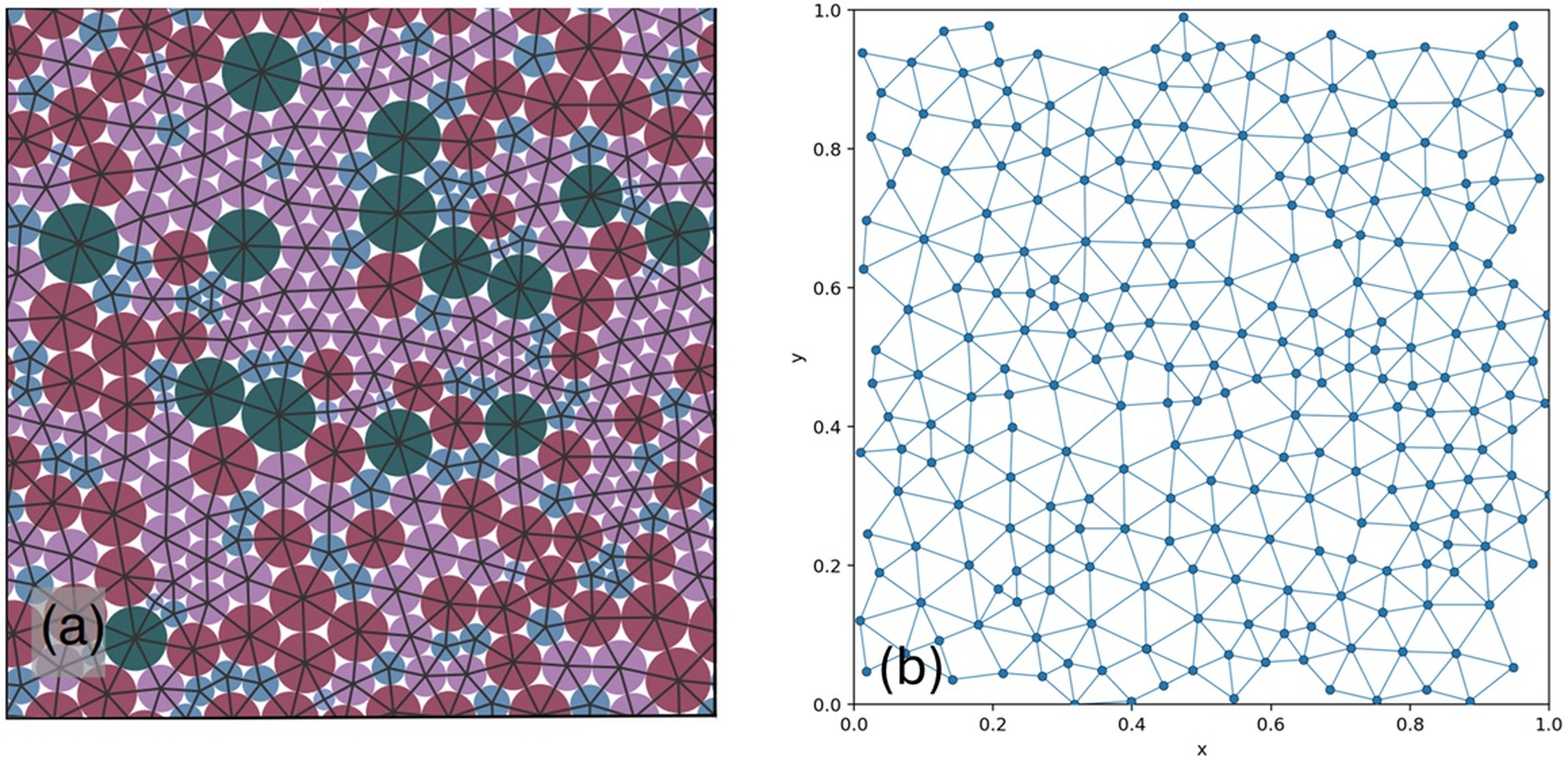}
    \caption{
(a) Numerical packing of polydisperse disks.
Panel (a) is reproduced from Bolton--Lum \textit{et al.}, \cite{Corwin}, with permission from the American Physical Society.
(b) Gabriel-graph constructed from the particle centers corresponding to the configuration in (a), used to define nearest-neighbor bonds for the computation of the inversion-symmetry order parameter $F_{\mathrm{IS}}$.
}
    \label{fig:2}
\end{figure}

We now test this hypothesis directly on the numerical data of Bolton-Lum et al. \cite{Corwin}. In particular, we analyze the ideal disk packing sample shown in Fig. \ref{fig:2}, and we compute the inversion-symmetry order parameter $F_{IS}$ as defined in \cite{Milkus2016}.

With reference to Fig. \ref{fig:1}, in the nonaffine lattice-dynamics framework of Milkus and Zaccone \cite{Milkus2016}, the microscopic driver of nonaffine motion is the force imbalance that appears when atoms are brought to their \emph{affine} positions under an imposed strain: in a locally centrosymmetric environment the nearest-neighbor forces cancel, whereas local inversion-symmetry breaking generically produces a nonzero net force that must be released by additional nonaffine displacements. In the harmonic approximation this imbalance can be written as $\mathbf{f}_i=\boldsymbol{\Xi}_i\,\gamma$, where $\gamma$ is the shear strain, and the vector $\boldsymbol{\Xi}_i$ defined as \cite{Lemaitre,Zaccone_book}:
\begin{equation}
\boldsymbol{\Xi}_i
=
-\kappa 
\sum_{j}
R_{ij}\mathbf{n}_{ij}\,
n_{ij}^{x}\,
n_{ij}^{y}
\end{equation}
where $\kappa$ is the spring constant of the particle-particle interaction, $R_{ij}$ is the scalar distance between particle $i$ and its nearest-neighbor $j$, and $\mathbf{n}_{ij}$ is the unit vector from $i$ to $j$.
The affine force vector $\boldsymbol{\Xi}_i$ encodes the degree of local inversion-symmetry breaking at site $i$ (it vanishes identically for locally centrosymmetric environments and is finite otherwise), thereby directly controlling the magnitude of nonaffine relaxations and the associated nonaffine softening of the elastic response. To quantify inversion-symmetry breaking as a structural indicator of shear rigidity, they define a normalized order parameter $F_{IS}$, based on the total affine force field magnitude $|\boldsymbol{\Xi}|^2$, namely
\begin{equation}
F_{\mathrm{IS}} \;=\; 1 - \frac{|\boldsymbol{\Xi}|^2}{\kappa^2  \sum_{ij}\!R_{ij}^2\left(n^{x}_{ij}n^{y}_{ij}\right)^{2}},
\label{eq:FIS_B5}
\end{equation}
which equals unity for perfect centrosymmetric lattices (where $|\boldsymbol{\Xi}|^2=0$) and tends to zero in the limit of maximally broken inversion symmetry, thus providing a direct structural predictor for the onset of nonaffine motions and the accompanying anomalies in shear elasticity and vibrational spectra. In the above formula,  \cite{Milkus2016,Zaccone_book}:
\begin{equation}
|\boldsymbol{\Xi}|^2
=
\kappa^2 
\sum_i
\sum_{\alpha \in \{x,y\}}
\left(
\sum_{j \in \mathrm{nn}(i)}
R_{ij}\,n_{ij}^{\alpha}\,
n_{ij}^{x}\,
n_{ij}^{y}
\right)^2 .
\end{equation}

In the above expressions, all summations over nearest neighbors are performed as directed sums $\sum_i\sum_{j\in \mathrm{nn}(i)}$, such that each bond contributes independently to the force imbalance of both particles it connects. In other words, each bond must contribute separately to each endpoint of the Gabriel graph, which is consistent with $F_{IS}$ being a network-averaged measure of local inversion symmetry directly interpretable as “how centrosymmetric the force environment of a typical node (atom $i$) is”.

Although the order parameter $F_{\mathrm{IS}}$ is constructed from the affine force field $\boldsymbol{\Xi}$, it is determined entirely by the static contact geometry and does not require evaluation of the dynamical (Hessian) matrix or elastic response. Therefore, the observed correlation between $F_{\mathrm{IS}}$ and the shear modulus is predictive rather than tautological, as $G$ depends additionally the Hessian and its exact calculation is much more demanding.

To identify nearest-neighbor pairs for the evaluation of the inversion-symmetry order parameter $F_{\mathrm{IS}}$, we first constructed the Delaunay triangulation of the particle centers from Fig. \ref{fig:2}(a) retained within the observation window. The corresponding Gabriel graph, shown in Fig. \ref{fig:2}(b) was then obtained by retaining only those Delaunay edges for which the closed disk having the edge as diameter contains no other particle centers. This procedure yields a parameter-free definition of nearest neighbors that is robust to local density fluctuations and does not rely on an arbitrary distance cutoff. The order parameter $F_{\mathrm{IS}}$ was subsequently evaluated as a bond-averaged quantity over the resulting set of Gabriel-connected particle pairs, using the pairwise separation $R_{ij}$ for each contributing bond.

From the numerical configuration comprising $288$ disk centers in Fig. \ref{fig:2}(a), particles whose centers lay outside the cropped observation window were excluded from the analysis. Specifically, a particle $i$ with coordinates $(x_i,y_i)$ was retained only if $0 \le x_i \le 1$ and $0 \le y_i \le 1$; all others were discarded. This criterion removed $33$ particles located near the boundaries, corresponding to an excluded fraction of approximately $11.5\%$ of the total population, leaving $N=255$ particles for the computation of structural observables. No additional distance-based or neighbor-based buffer was imposed: the exclusion was determined solely by the particle-center coordinates. The inversion-symmetry order parameter $F_{\mathrm{IS}}$ was then evaluated using the resulting bulk configuration, ensuring that all contributing interparticle bonds were fully contained within the observation window and free of boundary-induced incompleteness.

From the computation, we finally obtain the following estimate of $F_{IS}$ for the ultrastable ideal disk packing of Bolton-Lum et al. \cite{Corwin}:
\begin{equation}
    F_{IS}=0.84522
\end{equation}
which is indeed very close to the ideal defect-free centrosymmetric crystal limit, $F_{IS}=1$ \cite{Milkus2016}.

For completeness, we also compute the centrosymmetry parameter of Kelchner, Plimpton, and Hamilton \cite{Kelchner1998}, defined as:
\begin{equation}
CS
=
\sum_{i=1}^{N/2}
\left|
\mathbf{R}_i
+
\mathbf{R}_{i+N/2}
\right|^2 .
\end{equation}
Here the $N$ nearest neighbors of a given particle are first identified, and
$\mathbf{R}_i$ and $\mathbf{R}_{i+N/2}$ denote the vectors from the central
particle to a particular pair of nearest neighbors. Among the $N(N-1)/2$
possible neighbor pairs, the quantity
$\lvert \mathbf{R}_i + \mathbf{R}_j \rvert^2$ is evaluated for each pair, and the
$N/2$ smallest values are selected for the sum. These pairs typically correspond
to neighbors located in approximately opposite directions with respect to the
central particle, motivating the $i+N/2$ notation. The parameter $N$ is a user-defined
even integer chosen to match the coordination number of the local environment.
In a perfectly centrosymmetric environment, opposite neighbor vectors cancel
pairwise and $CS=0$, while local distortions and defects lead to finite values of
$CS$.

Using the same Gabriel graph employed for the computation of the inversion-symmetry order parameter $F_{\mathrm{IS}}$ above, we also evaluated the centrosymmetry parameter $CS$ for each particle. Nearest neighbors were identified from the Gabriel-graph connectivity, and the centrosymmetry parameter was computed using $N=6$, appropriate for two-dimensional ideal disk packings. Averaging over all particles within the bulk region, we obtain a mean value $\langle CS \rangle \simeq 4.6\times10^{-3}$, with a median of comparable magnitude. 
To facilitate comparison with previous work, we also consider a normalized
centrosymmetry parameter $CS^\ast = CS / R_0^2$, where $R_0$ is the mean
nearest-neighbor distance. Using the same Gabriel-defined neighbor network and
bulk particle set as for the inversion-symmetry order parameter $F_{\mathrm{IS}}$,
we find $\langle CS \rangle \simeq 4.6\times10^{-3}$ and $R_0 \simeq 7.1\times10^{-2}$,
yielding a normalized value $\langle CS^\ast \rangle \simeq 0.92$. This magnitude
is comparable to values reported for weakly distorted crystalline environments
and is significantly smaller than those associated with strong defects such as
dislocation cores or free surfaces. The result is consistent with the large value
of $F_{\mathrm{IS}}$ obtained for the same configuration above, indicating that local
environments remain close to centrosymmetric despite geometric disorder.

In particular, the magnitude of the normalized centrosymmetry parameter obtained here,
$\langle CS^\ast \rangle \simeq 0.9$, is comparable to values reported for
weakly distorted crystalline environments subject to thermal fluctuations
or elastic strain, for which $CS^\ast \sim 0.1{-}1$
\cite{Kelchner1998,Stukowski2010,Shimizu2007}. These values are significantly
smaller than those associated with stacking faults or twin boundaries
($CS^\ast \sim 2{-}6$) and are more than an order of magnitude smaller than
those characteristic of dislocation cores or free surfaces
($CS^\ast \gtrsim 5$) \cite{Kelchner1998,Li2003}. The observed high centrosymmetry
is therefore consistent with locally crystalline-like environments exhibiting
geometric disorder but lacking strong topological-like defects \cite{BaggioliPRL,Landry}.

The central result of this work is the demonstration that ideal two-dimensional
disk packings and ultrastable glasses possess an exceptionally high degree of
local inversion symmetry, quantitatively captured by the inversion-symmetry
order parameter $F_{\mathrm{IS}}$. The measured value
$F_{\mathrm{IS}}=0.84522$ is remarkably close to the crystalline limit
$F_{\mathrm{IS}}=1$ and stands in sharp contrast to the much smaller values
($F_{\mathrm{IS}}\sim0.3-0.5$) characteristic of ordinary jammed packings and
structural glasses. This finding provides a direct and physically transparent
structural explanation for the long-standing puzzle of why ideal glasses exhibit
crystal-like mechanical and vibrational properties despite the absence of
long-range order.
The large value of
$F_{\mathrm{IS}}$ measured here naturally accounts for both the
anomalously high rigidity and the crystal-like vibrational spectrum reported for
ideal disk packings, without invoking hidden crystalline order or exotic
mechanisms.

The present analysis also clarifies the limitations of traditional structural
metrics. Bond-orientational order parameters, while sensitive to angular
correlations, are fundamentally blind to inversion symmetry and therefore fail
to correlate with mechanical stiffness or vibrational anomalies in disordered
solids. In contrast, $F_{\mathrm{IS}}$ directly probes the symmetry properties of
local force transmission and thus provides a robust structural predictor of
elastic and vibrational behavior. The consistency between the large
$F_{\mathrm{IS}}$ and the small, crystal-like value of the normalized
centrosymmetry parameter $CS^\ast$ further reinforces this interpretation.

More broadly, these results suggest that local centrosymmetry constitutes a key
organizing principle for amorphous solids at the edge between glassy and
crystalline behavior. Ideal glasses and ideal packings emerge as systems that
remain topologically disordered yet are statistically optimized to maximize
local inversion symmetry, thereby suppressing nonaffine softening. This insight
offers a unifying geometric framework to understand ultrastability, hyperuniform
mechanical response, and the disappearance of low-frequency vibrational
excitations. Beyond two-dimensional disk packings, the same concepts are
expected to apply to three-dimensional glasses, jammed materials (including frictional suspensions and shear-thickening systems \cite{DAmico2025}), and metallic
glasses \cite{Wei-Hua2025}, opening the door to symmetry-based design principles for mechanically
robust amorphous materials.

\subsection*{Acknowledgments} 
A.Z. gratefully acknowledges funding from the European Union through Horizon Europe ERC Grant number: 101043968 ``Multimech'', from US Army Research Office through contract nr. W911NF-22-2-0256, and from the Nieders{\"a}chsische Akademie der Wissenschaften zu G{\"o}ttingen in the frame of the Gauss Professorship program. 
\bibliographystyle{apsrev4-2}

\bibliography{refs}

@article{Amelia2,
author = "Liu, Amelia C. Y. and Pham, Huyen and Bera, Arabinda and Petersen, Timothy C. and Sirk, Timothy W. and Mudie, Stephen T. and Tabor, Rico F. and Nunez-Iglesias, Juan and Zaccone, Alessio and Baggioli, Matteo",
title = "{Geometric indicators of local plasticity in glasses measured by scanning small-beam diffraction}",
journal = "Acta Crystallographica Section A",
year = "2026",
volume = "82",
number = "1",
pages = "4--17",
month = "Jan",
doi = {10.1107/S2053273325009775},
url = {https://doi.org/10.1107/S2053273325009775},
}

@article{
Amelia,
author = {Amelia C. Y. Liu  and Espen D. Bøjesen  and Rico F. Tabor  and Stephen T. Mudie  and Alessio Zaccone  and Peter Harrowell  and Timothy C. Petersen },
title = {Local symmetry predictors of mechanical stability in glasses},
journal = {Science Advances},
volume = {8},
number = {11},
pages = {eabn0681},
year = {2022},
doi = {10.1126/sciadv.abn0681},
URL = {https://www.science.org/doi/abs/10.1126/sciadv.abn0681}}

@article{Corwin,
  title = {Ideal Glass and Ideal Disk Packing in Two Dimensions},
  author = {Bolton-Lum, Viola M. and Dennis, R. Cameron and Morse, Peter K. and Corwin, Eric I.},
  journal = {Phys. Rev. Lett.},
  volume = {136},
  issue = {5},
  pages = {058201},
  numpages = {7},
  year = {2026},
  month = {Feb},
  publisher = {American Physical Society},
  doi = {10.1103/vldy-r77w},
  url = {https://link.aps.org/doi/10.1103/vldy-r77w}
}

@book{Zaccone_book,
  title={Theory of Disordered Solids},
  author={Zaccone, A.},
  year={2023},
  place={Cham},
 doi={10.1007/978-3-031-24706-4},
  publisher={Springer}
}

@article{Zaccone_rev,
    author = {Zaccone, Alessio},
    title = {Complete mathematical theory of the jamming transition: A perspective},
    journal = {Journal of Applied Physics},
    volume = {137},
    number = {5},
    pages = {050901},
    year = {2025},
    month = {02},
    issn = {0021-8979},
    doi = {10.1063/5.0245684},
    url = {https://doi.org/10.1063/5.0245684}
}

@article{
McKenna,
author = {Heedong Yoon  and Gregory B. McKenna },
title = {Testing the paradigm of an ideal glass transition: Dynamics of an ultrastable polymeric glass},
journal = {Science Advances},
volume = {4},
number = {12},
pages = {eaau5423},
year = {2018},
doi = {10.1126/sciadv.aau5423},
URL = {https://www.science.org/doi/abs/10.1126/sciadv.aau5423},
eprint = {https://www.science.org/doi/pdf/10.1126/sciadv.aau5423}}

@article{torquato,
  title = {Local density fluctuations, hyperuniformity, and order metrics},
  author = {Torquato, Salvatore and Stillinger, Frank H.},
  journal = {Phys. Rev. E},
  volume = {68},
  issue = {4},
  pages = {041113},
  numpages = {25},
  year = {2003},
  month = {Oct},
  publisher = {American Physical Society},
  doi = {10.1103/PhysRevE.68.041113},
  url = {https://link.aps.org/doi/10.1103/PhysRevE.68.041113}
}

@article{ronchetti,
  title = {Bond-orientational order in liquids and glasses},
  author = {Steinhardt, Paul J. and Nelson, David R. and Ronchetti, Marco},
  journal = {Phys. Rev. B},
  volume = {28},
  issue = {2},
  pages = {784--805},
  numpages = {0},
  year = {1983},
  month = {Jul},
  publisher = {American Physical Society},
  doi = {10.1103/PhysRevB.28.784},
  url = {https://link.aps.org/doi/10.1103/PhysRevB.28.784}
}

@article{torquato_rev,
  title = {Jammed hard-particle packings: From Kepler to Bernal and beyond},
  author = {Torquato, S. and Stillinger, F. H.},
  journal = {Rev. Mod. Phys.},
  volume = {82},
  issue = {3},
  pages = {2633--2672},
  numpages = {0},
  year = {2010},
  month = {Sep},
  publisher = {American Physical Society},
  doi = {10.1103/RevModPhys.82.2633},
  url = {https://link.aps.org/doi/10.1103/RevModPhys.82.2633}
}

@article{OHern,
  title        = {Jamming at zero temperature and zero applied stress: The epitome of disorder},
  author       = {O'Hern, Corey S. and Silbert, Leonardo E. and Liu, Andrea J. and Nagel, Sidney R.},
  journal      = {Physical Review E},
  volume       = {68},
  number       = {1},
  pages        = {011306},
  year         = {2003},
  doi          = {10.1103/PhysRevE.68.011306}
}

@article{Taiki,
  title = {Generating Ultrastable Glasses by Homogenizing the Local Virial Stress},
  author = {Leoni, Fabio and Russo, John and Sciortino, Francesco and Yanagishima, Taiki},
  journal = {Phys. Rev. Lett.},
  volume = {134},
  issue = {12},
  pages = {128201},
  numpages = {7},
  year = {2025},
  month = {Mar},
  publisher = {American Physical Society},
  doi = {10.1103/PhysRevLett.134.128201},
  url = {https://link.aps.org/doi/10.1103/PhysRevLett.134.128201}
}

@Article{Schlegel2016,
author={Schlegel, M.
and Brujic, J.
and Terentjev, E. M.
and Zaccone, A.},
title={Local structure controls the nonaffine shear and bulk moduli of disordered solids},
journal={Scientific Reports},
year={2016},
month={Jan},
day={06},
volume={6},
number={1},
pages={18724},
issn={2045-2322},
doi={10.1038/srep18724},
url={https://doi.org/10.1038/srep18724}
}

@article{JAP,
    author = {Zaccone, Alessio and Terentjev, Eugene M.},
    title = {Short-range correlations control the G/K and Poisson ratios of amorphous solids and metallic glasses},
    journal = {Journal of Applied Physics},
    volume = {115},
    number = {3},
    pages = {033510},
    year = {2014},
    month = {01},
    issn = {0021-8979},
    doi = {10.1063/1.4862403},
    url = {https://doi.org/10.1063/1.4862403}
}

@article{Stukowski2010,
  author  = {Stukowski, Alexander},
  title   = {Visualization and analysis of atomistic simulation data with {OVITO}},
  journal = {Modelling and Simulation in Materials Science and Engineering},
  volume  = {18},
  number  = {1},
  pages   = {015012},
  year    = {2010},
  doi     = {10.1088/0965-0393/18/1/015012}
}

@article{Shimizu2007,
  author  = {Shimizu, F. and Ogata, S. and Li, J.},
  title   = {Theory of shear banding in metallic glasses and molecular dynamics calculations},
  journal = {Acta Materialia},
  volume  = {55},
  number  = {18},
  pages   = {6096--6106},
  year    = {2007},
  doi     = {10.1016/j.actamat.2007.07.012}
}

@article{Zaccone2022,
  title = {Explicit Analytical Solution for Random Close Packing in $d=2$ and $d=3$},
  author = {Zaccone, Alessio},
  journal = {Phys. Rev. Lett.},
  volume = {128},
  issue = {2},
  pages = {028002},
  numpages = {5},
  year = {2022},
  month = {Jan},
  publisher = {American Physical Society},
  doi = {10.1103/PhysRevLett.128.028002},
  url = {https://link.aps.org/doi/10.1103/PhysRevLett.128.028002}
}

@article{Li2003,
  author  = {Li, Ju},
  title   = {Atomistic simulations of dislocation nucleation and propagation},
  journal = {Modelling and Simulation in Materials Science and Engineering},
  volume  = {11},
  number  = {2},
  pages   = {173--177},
  year    = {2003},
  doi     = {10.1088/0965-0393/11/2/304}
}

@article{Kelchner1998,
  author  = {Kelchner, C. L. and Plimpton, S. J. and Hamilton, J. C.},
  title   = {Dislocation nucleation and defect structure during surface indentation},
  journal = {Phys. Rev. B},
  volume  = {58},
  number  = {17},
  pages   = {11085--11088},
  year    = {1998},
  doi     = {10.1103/PhysRevB.58.11085}
}

@article{Landry,
  title = {Deformations, relaxation, and broken symmetries in liquids, solids, and glasses: A unified topological field theory},
  author = {Baggioli, Matteo and Landry, Michael and Zaccone, Alessio},
  journal = {Phys. Rev. E},
  volume = {105},
  issue = {2},
  pages = {024602},
  numpages = {22},
  year = {2022},
  month = {Feb},
  publisher = {American Physical Society},
  doi = {10.1103/PhysRevE.105.024602},
  url = {https://link.aps.org/doi/10.1103/PhysRevE.105.024602}
}

@article{BaggioliPRL,
  title = {Plasticity in Amorphous Solids Is Mediated by Topological Defects in the Displacement Field},
  author = {Baggioli, Matteo and Kriuchevskyi, Ivan and Sirk, Timothy W. and Zaccone, Alessio},
  journal = {Phys. Rev. Lett.},
  volume = {127},
  issue = {1},
  pages = {015501},
  numpages = {6},
  year = {2021},
  month = {Jul},
  publisher = {American Physical Society},
  doi = {10.1103/PhysRevLett.127.015501},
  url = {https://link.aps.org/doi/10.1103/PhysRevLett.127.015501}
}

@article{Plimpton,
  title = {Dislocation nucleation and defect structure during surface indentation},
  author = {Kelchner, Cynthia L. and Plimpton, S. J. and Hamilton, J. C.},
  journal = {Phys. Rev. B},
  volume = {58},
  issue = {17},
  pages = {11085--11088},
  numpages = {0},
  year = {1998},
  month = {Nov},
  publisher = {American Physical Society},
  doi = {10.1103/PhysRevB.58.11085},
  url = {https://link.aps.org/doi/10.1103/PhysRevB.58.11085}
}

@article{Zaccone2025_RCP2D,
  title        = {Analytical solution for the polydisperse random close packing problem in 2D},
  author       = {Zaccone, Alessio},
  journal      = {Powder Technology},
  volume       = {459},
  pages        = {121008},
  year         = {2025},
  doi          = {10.1016/j.powtec.2025.121008},
  note         = {Analytical expressions for maximal packing fractions of polydisperse hard disks up to \(\phi\gtrsim 0.96\) as a function of size distribution width},
}

@Article{Wang2025,
author={Wang, Yinqiao
and Qian, Zhuang
and Tong, Hua
and Tanaka, Hajime},
title={Hyperuniform disordered solids with crystal-like stability},
journal={Nature Communications},
year={2025},
month={Feb},
day={12},
volume={16},
number={1},
pages={1398},
issn={2041-1723},
doi={10.1038/s41467-025-56283-1},
url={https://doi.org/10.1038/s41467-025-56283-1}
}

@article{entropy,
  title = {Atomistic structural mechanism for the glass transition: Entropic contribution},
  author = {Han, Dong and Wei, Dan and Yang, Jie and Li, Hui-Ling and Jiang, Min-Qiang and Wang, Yun-Jiang and Dai, Lan-Hong and Zaccone, Alessio},
  journal = {Phys. Rev. B},
  volume = {101},
  issue = {1},
  pages = {014113},
  numpages = {8},
  year = {2020},
  month = {Jan},
  publisher = {American Physical Society},
  doi = {10.1103/PhysRevB.101.014113},
  url = {https://link.aps.org/doi/10.1103/PhysRevB.101.014113}
}

@article{
Ediger,
author = {Stephen F. Swallen  and Kenneth L. Kearns  and Marie K. Mapes  and Yong Seol Kim  and Robert J. McMahon  and M. D. Ediger  and Tian Wu  and Lian Yu  and Sushil Satija },
title = {Organic Glasses with Exceptional Thermodynamic and Kinetic Stability},
journal = {Science},
volume = {315},
number = {5810},
pages = {353-356},
year = {2007},
doi = {10.1126/science.1135795},
URL = {https://www.science.org/doi/abs/10.1126/science.1135795},
eprint = {https://www.science.org/doi/pdf/10.1126/science.1135795}}

@article{Milkus2016,
  title = {Local inversion-symmetry breaking controls the boson peak in glasses and crystals},
  author = {Milkus, R. and Zaccone, A.},
  journal = {Phys. Rev. B},
  volume = {93},
  issue = {9},
  pages = {094204},
  numpages = {10},
  year = {2016},
  month = {Mar},
  publisher = {American Physical Society},
  doi = {10.1103/PhysRevB.93.094204},
  url = {https://link.aps.org/doi/10.1103/PhysRevB.93.094204}
}

@article{ZacconePRL,
  title = {Disorder-Assisted Melting and the Glass Transition in Amorphous Solids},
  author = {Zaccone, Alessio and Terentjev, Eugene M.},
  journal = {Phys. Rev. Lett.},
  volume = {110},
  issue = {17},
  pages = {178002},
  numpages = {5},
  year = {2013},
  month = {Apr},
  publisher = {American Physical Society},
  doi = {10.1103/PhysRevLett.110.178002},
  url = {https://link.aps.org/doi/10.1103/PhysRevLett.110.178002}
}

@article{Zaccone2011,
  title = {Approximate analytical description of the nonaffine response of amorphous solids},
  author = {Zaccone, Alessio and Scossa-Romano, Enzo},
  journal = {Phys. Rev. B},
  volume = {83},
  issue = {18},
  pages = {184205},
  numpages = {6},
  year = {2011},
  month = {May},
  publisher = {American Physical Society},
  doi = {10.1103/PhysRevB.83.184205},
  url = {https://link.aps.org/doi/10.1103/PhysRevB.83.184205}
}

@Article{Lemaitre,
author={Lema{\^i}tre, Ana{\"e}l
and Maloney, Craig},
title={Sum Rules for the Quasi-Static and Visco-Elastic Response of Disordered Solids at Zero Temperature},
journal={Journal of Statistical Physics},
year={2006},
month={Apr},
day={18},
volume={123},
number={2},
pages={415},
issn={1572-9613},
doi={10.1007/s10955-005-9015-5},
url={https://doi.org/10.1007/s10955-005-9015-5}
}

\end{document}